\documentstyle[preprint,pra,aps]{revtex}
\begin{document}
\draft
\title{Thermodynamical analogues in quantum information theory}
\author{Daniel Rohrlich  \thanks{E-mail: \tt rohrlich@post.tau.ac.il}}
\address{School of Physics and Astronomy, Tel Aviv University,
Ramat Aviv 69978 Tel Aviv, Israel}
\date{printed \today}
\maketitle
\begin{abstract}
The first step in quantum information theory is the identification of
entanglement as a valuable resource.  The next step is learning how to 
exploit this resource efficiently.  We learn how to exploit entanglement 
efficiently by applying analogues of thermodynamical concepts.  These 
concepts include reversibility, entropy, and the distinction between
intensive and extensive quantities.  We discuss some of these analogues
and show how they lead to a measure of entanglement for pure states.  We
also ask whether these analogues are more than analogues, and note that, 
{\it locally}, entropy of entanglement is thermodynamical entropy.
\end{abstract}

\tightenlines
\def\half{{1\over 2}}
\def\u{\vert\uparrow\rangle}
\def\d{\vert\downarrow\rangle}
\def\ua{\vert\uparrow_A \rangle}
\def\da{\vert\downarrow_A \rangle}
\def\ub{\vert\uparrow_B \rangle}
\def\db{\vert\downarrow_B \rangle}
\def\uk{\vert\uparrow_K \rangle}
\def\dk{\vert\downarrow_K \rangle}
\def\l{\lambda}
\def\sig{{\bf \sigma}}
\def\Pa{\vert\Psi_\alpha \rangle}

\newpage
\section{Introduction}

     We are familiar with the distinction between ``pure" and ``applied" 
research.  In pure research, knowledge of the physical world is an end 
in itself, while in applied research, knowledge is only a means to an 
end.  Usually, this distinction is a meaningful one; we can easily find
examples of pure research that has no applications, and of applied research
that does not increase our knowledge.  But sometimes the distinction is 
not meaningful.  For example, what about Carnot's invention of the cycle
that bears his name---was it pure research or applied research?  In this
example, the distinction simply doesn't exist.  It doesn't exist, and 
not only because Carnot invented his cycle while considering a practical 
engineering problem.  It doesn't exist, because his ``applied" research
on the efficiency of heat engines was also essential ``pure" research.  
Indeed, if Carnot had never considered a practical engineering problem, 
he would not have started thinking about limits to efficiency, and he 
would not have discovered the second law of thermodynamics.

     A second example, quite analogous to the first, arises in quantum 
information theory.  Research on quantum entanglement led to quantum 
information theory only after physicists found {\it uses} for entanglement 
and thought about how to use entanglement efficiently.  Limits on the 
efficient use of entanglement are fundamental to quantum information 
theory; hence applied research on entanglement is also essential pure 
research in quantum information theory.  The striking analogy between 
the roles of efficiency in thermodynamics and in quantum information 
theory is the subject of the next section.  This analogy leads us to 
several other analogies between the two theories.  Sect. III shows how 
analogues of heat engines, entropy, the thermodynamic limit and the 
second law of thermodynamics appear in quantum information theory.  The 
analogies between thermodynamics and quantum information theory are so 
striking that we ask, in the concluding section, if they are more than 
analogies.  Is quantum information theory, after all, a branch of 
thermodynamics?

\section{Entanglement as a resource}

     One of the most salient facts about quantum information theory is 
the fact that it came about recently, and not 60 or more years ago.  On 
the one hand, it could not have come before the birth of quantum mechanics 
in 1926 and the identification of entanglement by Schr\"odinger \cite{sch}
and Einstein, Podolsky and Rosen (EPR) \cite{epr} in 1935.  On the other
hand, it could have, in principle, come soon after 1935.  Why didn't it? 
What happened instead?  

     In the decades following the EPR paper, most physicists ignored 
it.  Bell was one of the few who did not; he agreed with EPR that ``no 
reasonable theory" should allow such an unreasonable thing as quantum 
entanglement.  Bell \cite{bell} published his famous inequalities almost
30 years after the EPR paper, and another five years passed until 
Clauser, Horne, Shimony and Holt (CHSH) \cite{chsh} suggested testing 
Bell's inequalities experimentally.  The CHSH paper sparked intense 
interest in entanglement, and the many papers that followed it demonstrated
again and again, in experiment and in theory, that quantum mechanics is 
unreasonable---but true.  Thus entanglement became a wall for physicists
to beat their heads against.  How can the world be so unreasonable?  
Even physicists who were ready to stop beating their heads against the 
wall lacked a sense of direction.  If we cannot understand entanglement, 
at least we could try, for example, to measure it, to quantify it.  But 
various proposed measures of entanglement seemed equally plausible 
\cite{mea}.

      What provided the direction was not the attempt to understand 
entanglement, but attempts to use it.  Consider singlet pairs of spin-1/2
particles or photons shared by two remote observers, Alice and Bob; 
Alice has one particle in each pair, and Bob has the other.  Alice and Bob
can use these pairs in at least two ways.  They can construct unbreakable
codes, and they can teleport quantum states.  Quantum cryptography began
with a paper by Wiesner \cite{weisner} in 1983; in 1991, Ekert \cite{ekert}
applied entanglement to quantum cryptography.  Quantum teleportation 
appeared in a paper by Bennett et al. \cite{telep} in 1993.  Let us 
consider each of these applications in turn.  

     The use of entanglement in quantum cryptography is straightforward.
On each shared singlet pair, Alice and Bob measure polarization along
identical axes.  Since the pair is a singlet, Alice and Bob will always 
find the same (for photons) or opposite (for spin-1/2 particles) 
polarizations.  Either way, Alice and Bob can thus construct identical
sequences of binary data that only they know.  Suppose that Alice wants 
to send a coded message to Bob.  First, she translates her message into
binary.  For example, let the binary message be 1001011101110001010100010,
which contains 25 bits.  Next, she and Bob generate a shared random 
binary sequence of the same length.  Then Alice adds the two sequences,
in binary, and transmits the sum to Bob using any (public or private) 
channel.  Finally, Bob subtracts the random binary sequence from the sum
and recovers Alice's message.  Only Bob can read Alice's message, because
only Bob knows what to subtract. 

     Teleportation, as everyone knows, is Captain Kirk's way of getting 
around.  He would turn into a column of glimmering light and disappear,
only to reappear far away.  To explain teleportation, we approximate
Captain Kirk as a single spin-1/2 particle in an unknown state
$\vert K\rangle$: 
\begin{equation}
\vert K\rangle =a\uk +b\dk~~~~.
\end{equation}
($K$ stands for Kirk).  Suppose Bob wants to transmit the state $\vert K 
\rangle$.  Alice and Bob join to this state a singlet pair, such that the
overall state is
\begin{equation}
{1\over\sqrt{2}} (a\uk +b\dk ) \left[ \ua\db -\da\ub \right]~~~,
\label{abk}
\end{equation}
where $A$ and $B$ stand for Alice and Bob, respectively.  Next, Bob 
measures a nondegenerate operator with four eigenstates $\vert \Psi^{
(-)}\rangle$, $\vert \Psi^{(+)} \rangle$, $\vert\Phi^{(-)} \rangle$ and 
$\vert \Phi^{(+)} \rangle$:
\begin{eqnarray}
\vert \Psi^{(+)} \rangle &=& {1\over\sqrt{2}} \left[ \uk\db +\dk\ub \right]
~~~,\nonumber\\
\vert \Psi^{(-)} \rangle &=& {1\over\sqrt{2}} \left[ \uk\db -\dk\ub \right]
~~~,\nonumber\\
\vert \Phi^{(+)} \rangle &=& {1\over\sqrt{2}} \left[ \uk\ub +\dk\db \right]
~~~,\nonumber\\
\vert \Phi^{(-)} \rangle &=& {1\over\sqrt{2}} \left[ \uk\ub -\dk\db \right]
~~~~.
\end{eqnarray}
These states are called the Bell operator basis.  Rewriting Eq.\ 
(\ref{abk}) in this basis, we obtain
\begin{eqnarray}
\half \vert \Psi^{(+)} \rangle (a\ua &-& b\da )
+ \half \vert \Psi^{(-)} \rangle (a\ua +b\da )\nonumber\\
&-&\half\vert \Phi^{(+)} \rangle (a\da -b\ua )
-\half \vert \Phi^{(-)} \rangle (a\da +b \ua )
\end{eqnarray}
as the state of all three spins before Bob's measurement.  Now Bob's
measurement leaves the two spins at his end in one of the states of 
the Bell operator basis, and he knows which one from his measurement.  
Bob transmits this information---his two bits---to Alice; he can even
broadcast the information without knowing where Alice is.  When Alice
receives the two bits, she infers the state of her spin.  If the state 
is $a\ua -b\da$, she rotates her spin $\pi$ around the $z$-axis.  If 
it is $a\ua +b\da$, she does nothing.  If it is $a\da -b\ua$, she rotates
her spin $\pi$ around the $y$-axis.  If it is $a\da +b \ua$, she rotates
her spin $\pi$ around the $x$-axis.  In each case, the final state of 
her spin is a replica of Captain Kirk, $a\ua +b\da$.  No trace of 
Captain Kirk remains with Bob.
 
     Hence every singlet pair that Alice and Bob share is a valuable
resource, which Alice and Bob can use to encode one bit of a message 
or to teleport a single qubit.  Also, they can use any pair that is
related to a singlet pair by local unitary transformations.  We
call such a pair an {\it ebit}.  For these applications of entanglement,
only ebits will do; pairs in other entangled states would, in general, 
yield errors in transmission of a message or a qubit.  But if Alice and 
Bob share pairs in other entangled states, they can, with a certain 
probability, extract ebits from these pairs.  Here's how:  Suppose 
Alice and Bob share the entangled state $\Pa$:
\begin{equation}
\Pa =  \alpha \ua\ub
+ (1-\alpha^2 )^{1/2} \da\db
~~~,
\end{equation}
with $\alpha$ real and $\alpha >1/\sqrt{2}$.  Since $\alpha$ is too 
large, either Alice or Bob must reduce $\alpha$ by local filtering 
\cite{bbps}.  So let Alice, say, run her spin through a selective 
filter that never absorbs the state $\da$, but sometimes absorbs the 
state $\ua$.  Let $\vert 0\rangle$ denote the initial state of the filter
and define local filtering according to a unitary operator $U$ that sends
\begin{eqnarray}
\ua \vert 0\rangle &\rightarrow & x \ua \vert 0\rangle 
+y\ua \vert 1\rangle ~~~,\nonumber\\
\da \vert 0 \rangle &\rightarrow & \da \vert 0\rangle
~~~~.
\end{eqnarray}
Here $\vert 1\rangle$ represents the state of the filter if it absorbs
Alice's spin, and $\vert x\vert^2 +\vert y\vert^2 =1$.  After Alice
runs her spin through the filter, the combined state of the two spins 
and the filter is
\begin{equation}
\left[ x\alpha \ua\ub+ (1-\alpha^2 )^{1/2}
\da\db \right]\vert 0\rangle
+y \alpha\ua\ub \vert 1\rangle~~~~.\label{ep}
\end{equation}
Now Alice looks in the filter.  The chance is $\vert \alpha y\vert^2$ 
that she finds her spin there.  If she does not find her spin in the 
filter, however, she knows that the state of the two spins is given by 
the bracketed term in Eq.\ (\ref{ep}), up to normalization.  In 
particular, if we choose $x = (1-\alpha^2 )^{1/2} /\alpha$, the state of
the two spins will now be an ebit, suitable for coding and teleportation.  
So Alice has a chance $1- \vert \alpha y\vert^2 = 2(1 -\alpha^2)$ of 
producing an exact ebit.  Of course, she loses all the entanglement in 
$\Pa$ if the filter absorbs the $\ua$ state, but with a little luck, 
Alice and Bob can turn pairs in the state $\Pa$ into a valuable resource.  
(It follows that pairs in the state $\Pa$ are a valuable resource, too.)

     So we can extract singlets from other entangled pairs!  Immediately,
the question arises---just as it did for Carnot---whether local filtering
is the most efficient method for extracting singlets.  The answer is 
that, in general, it is not.  If Alice and Bob share many pairs in the 
state $\Pa$, they can do better than locally filter the pairs, one by 
one.  But to do better, Alice and Bob must apply {\it collective} 
operations to their entangled pairs---they must operate on many pairs 
together, and not one by one.  Suppose Alice and Bob share {\it two} 
pairs in the entangled state $\Pa$.  The state of two pairs is
\begin{equation}
\left[ \alpha \ua\ub
+ (1-\alpha^2 )^{1/2} \da\db \right]
\left[ \alpha \vert \uparrow_{A^\prime}\rangle 
\vert \uparrow_{B^\prime} \rangle
+ (1-\alpha^2 )^{1/2} \vert \downarrow_{A^\prime}
\rangle \vert \downarrow_{B^\prime}
\rangle \right]~~~,\label{pairs}
\end{equation}
where $A$ and $A^\prime$ refer to Alice's spins, and $B$ and $B^\prime$
refer to Bob's.  Expanding Eq.\ (\ref{pairs}), we obtain
\begin{eqnarray}
\alpha^2 \ua\ub \vert \uparrow_{A^\prime} \rangle 
\vert \uparrow_{B^\prime} \rangle &+& (1-\alpha^2 ) 
\da\db \vert \downarrow_{A^\prime} \rangle 
\vert \downarrow_{B^\prime} \rangle \nonumber\\
&+&\sqrt{2} \alpha (1-\alpha^2 )^{1/2}  \left[ {{\ua\ub
\vert \downarrow_{A^\prime} \rangle
\vert \downarrow_{B^\prime} \rangle 
+ \da\db
\vert \uparrow_{A^\prime} \rangle
\vert \uparrow_{B^\prime} \rangle } \over \sqrt{2}}\right]  \label{brac}
\end{eqnarray}
as the overall state.  Now let Bob measure $\sigma_z^B + \sigma_z^{B^
\prime}$, or Alice measure $\sigma_z^A + \sigma_z^{A^\prime}$.  (The result
is the same.)  With probability $2\alpha^2 (1-\alpha^2 )$ the result is
zero, and the state of the spins after the measurement is the bracketed
term in Eq.\ (\ref{brac}).  Now the bracketed term is an ebit; we can
define 
\begin{eqnarray}
\vert \Uparrow_A \rangle \equiv \ua
\vert\downarrow_{A^\prime}\rangle~~~~&,&~~~~\vert \Downarrow_A 
\rangle \equiv \da \vert \uparrow_{A^\prime}\rangle \nonumber\\
\vert \Downarrow_B \rangle \equiv -\ub
\vert \downarrow_{B^\prime}\rangle~~~~&,&~~~~\vert \Uparrow_B 
\rangle \equiv \db \vert \uparrow_{B^\prime}\rangle 
\end{eqnarray}
to write it explicitly as a singlet.

     For two pairs, this method is not more efficient than local 
filtering.  But it gets more efficient as Alice and Bob apply it 
collectively to more pairs at a time \cite{bbps}.  To apply the method 
to many pairs at a time, Alice and Bob first measure the $z$-component 
of total spin of the pairs.  Whatever result they get, their pairs are
left in a superposition of biorthogonal states with coefficients of
equal magnitude.  For example, let Alice and Bob have three distinct
pairs in the state $\vert\Psi_\alpha\rangle$.  Suppose Alice measures 
the $z$-component of total spin to be $1/2$; then the state of
their pairs is
\begin{eqnarray}
\ua\ub
\vert \uparrow_{A^\prime}\rangle
\vert \uparrow_{B^\prime}\rangle
\vert \downarrow_{A^{\prime\prime}}\rangle
\vert \downarrow_{B^{\prime\prime}}\rangle
&+&
\da\db
\vert \uparrow_{A^\prime}\rangle
\vert \uparrow_{B^\prime}\rangle
\vert \uparrow_{A^{\prime\prime}}\rangle
\vert \uparrow_{B^{\prime\prime}}\rangle
\nonumber\\
&&~~+
\ua\ub
\vert \downarrow_{A^\prime}\rangle
\vert \downarrow_{B^\prime}\rangle
\vert \uparrow_{A^{\prime\prime}}\rangle
\vert \uparrow_{B^{\prime\prime}}\rangle
~~~,\label{three}
\end{eqnarray}
up to normalization.  Each time they measure the $z$-component of total 
spin on a set of pairs, they get a state of this form.  The tensor 
product of such states is therefore also a state of this form.  For 
example, two groups of three spins, each in the state of Eq.\ 
(\ref{three}), yield a tensor product state having nine terms with
equal coefficents.  Alice and Bob can build such states, with various 
numbers of terms in the superposition, until they get a number of 
terms equal to, or nearly equal to, a power of 2.  A superposition 
with number of terms equal to $2^n$ is unitarily equivalent to $n$ 
ebits.  Thus Alice and Bob can apply local unitary operations to 
transform their superposition into ebits.  For example, Alice and Bob
could, by locally filtering a state having nine terms, reduce nine terms 
to eight terms.  Eight terms are unitarily equivalent to three ebits.  
Bennett, Bernstein, Popescu and Schumacher (BBPS) \cite{bbps} showed 
that Alice and Bob can obtain $n$ singlets from $k$ pairs of spins in 
the state $\Pa$, where the ratio $n/k$ approaches the limit
\begin{eqnarray}
\lim_{n,k\rightarrow \infty} {n\over k} &=& E(\Pa ) \nonumber\\
& =& -\alpha^2\log_2 \alpha^2 - (1-\alpha^2 ) \log_2 (1-\alpha^2 )~~~~.
\label{ee}
\end{eqnarray}
$E(\Pa )$ is called the {\it entropy of entanglement}; it is the Shannon
entropy of the squares of the coefficients of the Schmidt decomposition. 
It equals 1 if $\alpha =1/\sqrt {2}$ and equals 0 for a product state.

\section{Thermodynamical analogues}          

     Applications of entanglement naturally raise the question of 
efficiency.  We have two methods to extract singlets from generic 
entanglement; we know they are not equally efficient.  Are there more 
efficient methods?  Is there a maximally efficient method?  These
questions are analogous to the questions that Carnot asked.  We will 
now see also that the answers are analogous to his answers.

     First, the answer to the thermodynamical question involves a 
principle---the principle that there is no way to build a {\it 
perpetuum mobile}, i.e. to build a machine that works for free, 
without changing its environment.  This principle is the second law of 
thermodynamics.  For entanglement, there is an analogous principle:  
local operations cannot increase the entanglement between remote 
systems \cite{m,pr}.  Measurements, local unitary operations, and 
additional unentangled systems cannot increase the entanglement between 
Alice's and Bob's systems; neither can classical communication (i.e. 
communication that does not involve entanglement) between Alice and 
Bob.  We can accept this principle as an axiom, or we can prove it
as a theorem of quantum mechanics \cite{pr}.  It is analogous to the
second law also in that it is a statistical law.  The method of local 
filtering, for example, can increase the entanglement between the
systems held by Alice and Bob, but on average it decreases the 
entanglement.

     Second, Carnot had the insight to focus on {\it reversible} 
transformations.  Consider two reversible heat engines; suppose that
both absorb heat $Q_1$ at $T_1$ and expel heat $Q_2$ at $T_2$, but one
does work $W$, and the other does work $W^\prime >W$, per cycle.  The
first engine, if run in reverse, is a refrigerator---{\it absorbs} heat 
$Q_2$  at $T_2$ and {\it expels} heat $Q_1$ at $T_1$---and requires
only work  $W$ per cycle.  Thus the two engines together could provide 
$W^\prime  -W$ in work per cycle without changing their environment.  
Such a  conclusion contradicts the second law of thermodynamics, so 
both engines must do the same  work: $W=W^\prime$.

     Are the collective operations of BBPS reversible?  Alice and Bob
{\it can} turn shared singlets into shared pairs in the state $\Pa$.
Alice, say, can prepare pairs in the state $\Pa$ in her laboratory, and 
then teleport one spin out of each entangled pair to Bob.  However, Alice
then uses up one singlet pair for every spin that she teleports to Bob,
so Alice and Bob use up $k$ shared singlets to produce $k$ shared pairs 
in the state $\Pa$, whereas from $k$ pairs in the state $\Pa$ they can 
recover only $n<k$ singlet pairs.  So this is not an efficient way to 
produce pairs in the state $\Pa$.  But Alice can teleport the pairs more
efficiently using a method called {\it quantum data compression} \cite{qdc}. 
The idea behind quantum data compression is as follows.  Alice has to 
teleport $k$ spins, i.e. a state in a $2^k$-dimensional Hilbert space.  
But the effective dimension of the Hilbert space is much smaller than 
$2^k$, because the $k$ spins have a common bias.  In the state $\Pa$ with 
$\alpha>1$, $\u$ is more likely than $\d$, so a sequence with every
spin in the state $\u$ is much more likely than a sequence with every spin
in the state $\d$; still more likely are sequences with most, but not all,
spins in the state $\u$.  In fact, the effective dimension of the Hilbert
space approaches $2^n$, rather than $2^k$; that is, Alice can actually 
teleport the $k$ spins to Bob without using more than the $n$ singlets 
that she and Bob can obtain from $k$ pairs in the state $\Pa$.  Hence
the extraction of singlets from pairs in the state $\Pa$ is reversible.  
It is not reversible for finite $k$ and $n$, in general, but it is
reversible as the number of systems approaches infinity, just as heat
engines are reversible only in the thermodynamical limit.

     Now suppose that Alice and Bob share $k$ pairs of systems in an 
entangled state $\Pa$, which they transform into $n$ singlets, using 
the method of BBPS.  Did they use the most efficient method possible?  
That is, could Alice and Bob apply a more efficient method, using only 
local operations and classical communication, to obtain a greater number 
$n^\prime >n$ of singlets from the same number $k$ of initial pairs?
The answer is that they cannot.  For if it were possible to transform 
$k$ of the initial pairs into $n^\prime$ singlets by a different method,
Alice and Bob could then reverse the BBPS operations on $n$ of the
singlets and transform them into $k$ pairs in the entangled state $\Pa$. 
They could then obtain $n^\prime -n$ entangled pairs only using local
operations and classical communication, contradicting the general 
principle that local operations cannot increase entanglement.  Hence
$n^\prime =n$; the BBPS method is maximally efficient (in the limit
$k,n\rightarrow\infty$).

     As a byproduct, this proof gives us a measure of the entanglement
of the state $\Pa$.  The $k$ systems in the state $\Pa$ have the same 
entanglement as $n$ singlet pairs.  Thus the measure of entanglement for
$k$ pairs in the state $\Pa$ must equal the measure of entanglement for 
$n$ singlets.  At first, it might seem that we could assign an arbitrary
measure of entanglement, such as $n$, $n^2$ and $e^n$, to $n$ singlets.
But actually, the measure must be proportional to $n$, because the 
BBPS collective operations are reversible only when the number of systems
becomes arbitrarily large.  (The ratio $n/k$ nearly always tends to an 
irrational number, and if the number is irrational, we can never reversibly
extract $n$ singlets from a finite number $k$ of systems.)  Reversibility
requires us to go to the limit of infinite $n$, and for infinite $n$ there
is no way to define total entanglement.  We can only define entanglement
{\it per system}.  Here too, we find a thermodynamical analogue:  the 
thermodynamic limit requires us to define intensive quantities.  Likewise,
the measure of entanglement must be intensive, i.e. the measure of 
entanglement of $n$ singlets must be proportional to $n$.  It follows 
that the measure of entanglement for pure states is unique (up to a 
constant factor).  Since the measure of entanglement of $k$ systems 
$\Pa$ approaches the measure of entanglement of $n$ pairs in a singlet 
state, and since the measure is intensive, we have $kE(\Pa ) =n$, where 
$E$ now denotes the measure, and the measure of entanglement of a singlet 
state is 1.  Thus
\begin{equation}
E(\Pa ) =\lim_{n,k \rightarrow \infty}
{n\over k} ~~~~.
\end{equation}
This limit indeed equals the entropy of entanglement of $\Pa$, Eq.\
(\ref{ee}); so the measure of entanglement of $\Pa$ must equal its entropy
of entanglement, up to a conventional proportionality constant---measuring
the entanglement of a singlet pair or ebit---that we set it to 1. 

\section{Beyond analogues}

     Thermodynamical analogues are powerful tools for quantum
information theory.  However, so far they remain mere analogues. Feynman
\cite{feyn} constructed an amusing analogue of the Carnot cycle to prove
that gravitational potential energy near the surface of the earth is the
product of weight and height.  But his proof does not make gravity a branch
of thermodynamics!  So thermodynamical analogues in quantum information 
theory are not necessarily more than analogues. In particular, the previous
section did not mention temperature, heat or heat baths in the context of
entanglement.  We did not need to ask how much work, if any, the BBPS 
method entails.  So far, we have no reason to consider quantum information
theory a branch of thermodynamics.  I claim, however, that it is.

     In a Carnot cycle, the entropy of the heat engine changes twice 
per cycle.  The heat engine absorbs entropy at the high temperature 
and releases entropy at the low temperature.  At first glance, the 
flow of entropy in the Carnot cycle seems not to fit the thermodynamical
analogues:  entropy changes in the Carnot cycle, while BBPS collective 
operations conserve the entropy of entanglement.  A closer look,
however, reveals yet another analogue.  The heat engine indeed absorbs
and releases entropy, but so does the environment, such that the total
entropy is unchanged.  Otherwise, the heat engine would not be reversible. 
Likewise, the BBPS operations would not be reversible if they did not
conserve entropy of entanglement.  We can therefore guess that the
systems shared by Alice and Bob are the analogue, not of the heat engine,
but of the heat engine plus environment.

     How does this analogy work?  Initially, say, Alice and Bob share
$k$ pairs of spins in an entangled state $\Pa$.  From these $k$ pairs
they extract, by the BBPS method, $n$ ebits, with $n<k$.  Aside from the
$n$ ebits, then, there remain $k-n$ spins; by conservation of entropy of
entanglement, these $k-n$ spins contain zero entanglement.  Thus, the
extraction of ebits suggests a process in which an ensemble of spins at
thermodynamical equilibrium divides into two ensembles, with differing 
average entropy per spin.  If we now visit, say, Alice's laboratory, and 
forget about Bob, we cannot help but describe the process as heating and
cooling of two subensembles: the $k-n$ spins are in a pure state, hence 
they are ``cold", while the $n$ ebits are in state of maximum entropy, 
hence they are ``hot".  The inverse of the BBPS method, in which $n$ 
ebits are combined with $k-n$ pairs of spins (shared by Alice and Bob) 
in product states, to yield $k$ equally entangled pairs, is a process 
in which two ensembles at different temperatures are reversibly brought 
to the same temperature.

     We might conclude, then, that the BBPS method consumes work---precisely
the amount of work required to separate the ensemble at thermodynamical
equilibrium reversibly into ``hot" and ``cold" ensembles---and that the 
inverse of the BBPS method produces work---precisely the amount of work 
produced when the ``hot" and ``cold" ensembles come reversibly into 
thermodynamical equilibrium.  However, this conclusion is premature.
The reason is that our use of the word ``reversibility" in the context
of entanglement does not quite match its usage in thermodynamics.  A
local increase of entropy in Alice's laboratory, or in Bob's, may conserve
the entanglement of the pairs they share, but it is not thermodynamically
reversible.  We must check whether the collective operations of the BBPS
method have {\it local} thermodyamical effects that we have not taken 
into account \cite{r}.  We have not done so here, because we have treated
collective operations abstractly, without considering their physical 
implementation.  But we {\it can} already conclude that entropy of 
entanglement is more than an analogue of thermodynamical entropy; 
locally, it {\it is} thermodynamical entropy. 

     It may seem paradoxical that Alice's spins can have entropy and
temperature (if we forget Bob) yet they belong to a pure state (if we
remember Bob).  But such is the nature of entanglement.  Consider a
closed system comprising a measuring device, a measured system, and an
environment, in an initial pure state.  After the measurement, which
entangles the measuring device and the measured system, decoherence sets
in.  The process of decoherence does not change the fact that the closed
system is in a pure state; it merely makes the fact irrelevant, for 
all practical purposes, because no one can keep track of it.  Furthermore,
even though the system as a whole is in a pure state, its entangled
subsystems are not.  The objective---not subjective---entropy of the
subsystems derives from entanglement \cite{yakir}.  In just this sense,
Alice's and Bob's spins have objective entropy.


\begin{references}

\bibitem{sch} E. Schr\"odinger, {\it Proc. Camb. Phil. Soc.} {\bf 31},
555 (1935).

\bibitem{epr} A. Einstein, B. Podolsky and N. Rosen, {\it Phys. Rev.} 
{\bf 47}, 777 (1935).

\bibitem{bell} J. S. Bell, {\it Physics} {\bf 1}, 195 (1964).

\bibitem{chsh} J. F. Clauser, M. A. Horne, A. Shimony, and R. A. Holt, 
{\it Phys. Rev. Lett.} {\bf 23}, 880 (1969).

\bibitem{mea} For a review, see A. Shimony, in {\it The Dilemma of 
Einstein, Podolsky and Rosen -- 60 Years Later} (Annals of the Israel 
Physical Society, {\bf 12}), A. Mann and M. Revzen, eds., Institute of 
Physics Publishing (1996).

\bibitem{weisner} S. Weisner, {\it Sigact News} {\bf 15}, 78 (1983).

\bibitem{ekert} A. K. Ekert, {\it Phys. Rev. Lett.} {\bf 67}, 661 (1991).

\bibitem{telep} C. H. Bennett, G. Brassard, C. Cr\'epeau, R. Jozsa, 
A. Peres and W. K. Wootters, {\it Phys. Rev. Lett.} {\bf 70}, 1895 (1993).

\bibitem{bbps} C. H. Bennett, H. J. Bernstein, S. Popescu and 
B. Schumacher, {\it Phys. Rev.} {\bf A53}, 2046 (1996).

\bibitem{m} C. H. Bennett, G. Brassard, S. Popescu, B. Schumacher, J. A.
Smolin and W. K. Wootters, {\it Phys. Rev. Lett.} {\bf 76}, 722 (1996).

\bibitem{pr} S. Popescu and D. Rohrlich, {\it Phys. Rev.} {\bf A56}, 
Rapid Comm. R3319 (1997);  S. Popescu and D. Rohrlich, The joy of 
entanglement, in {\it Introduction to Quantum Computation and 
Information}; eds. H.-K. Lo, S. Popescu, and T. P. Spiller (Singapore:  
World Scientific), 1998, pp. 29-48.

\bibitem{qdc} R. Jozsa and B. Schumacher, {\it J. Mod. Optics} {\bf 41}, 
2343 (1994); B. Schumacher, {\it Phys. Rev.} {\bf A51}, 2738 (1995).

\bibitem{feyn} R. P. Feynman, R. B. Leighton and M. Sands, {\it The
Feynman Lectures on Physics, I} (New York:  Addison-Wesley Pub. Co.), 
1963, Sect. 4-2.

\bibitem{r} D. Rohrlich, in preparation.

\bibitem{yakir} Y. Aharonov, personal communication.

\end{references}
\end{document}